\documentclass[twocolumn]{autart} 

\usepackage{graphicx}      
\usepackage{color}
\usepackage{comment}
\usepackage{amsmath} 
\usepackage{amssymb}  
\newtheorem{proposition}{Proposition}
\newtheorem{lemma}{Lemma}
\newtheorem{theorem}{Theorem}

\newtheorem{corollary}{Corollary}

\newtheorem{definition}{Definition}

\begin{document}
\begin{frontmatter}

\title{A Topological Approach for Computing  Supremal Sublanguages for Some Language Equations in Supervisory Control Theory\thanksref{footnoteinfo}} 

\thanks[footnoteinfo]{The research of the project was supported by Ministry of Education, Singapore, under grant AcRF TIER 1-2018-T1-001-245 (RG 91/18).}

\author[First]{Liyong Lin}, 
\author[First]{Rong Su} 

\address[First]{School of Electrical and Electronic Engineering, Nanyang Technological University, Singapore (e-mail: \{liyong.lin, rsu\}@ntu.edu.sg).}

\begin{abstract}                
In this paper, we shall present a topological approach for the computation of some supremal sublanguages, often specified by language equations,   which arise from the study of  the supervisory control theory. The basic idea is to identify the solutions of the language equations as open sets for some (semi)-topologies. Then, the supremal sublanguages naturally correspond to the supremal open subsets, i.e., the interiors. This provides an elementary and uniform approach for computing various supremal sublanguages encountered in the supervisory control theory and is closely related to a theory of approximation, known as the rough set theory, in artificial intelligence. 
\end{abstract}


\end{frontmatter}


                
\section{Introduction}
The supervisory control theory (SCT)  provides a rigorous framework for the automatic synthesis of correct-by-construction controllers. Given the models of the plant and  the specification, the theory provides tools for the automatic synthesis of a supervisor, i.e., a discrete-event controller, such that the closed-loop system is safe, i.e., bad state is never reachable, and non-blocking, i.e., goal state is always reachable~\cite{WC18}. 

The foundation of SCT is largely built based on the theory of  formal languages and automata~\cite{HU79}. In fact, quite many important properties such as controllability, normality, $L$-closure and so on~\cite{WC18}, which are used for characterizing the existence of a supervisor, can be naturally  specified with  language equations, by using the operations of prefix-closure, concatenation, intersection, projection, inverse projection and so on. The computation of the supremal sublanguages specified by
these language equations is a central task in supervisor synthesis~\cite{WC18}. 

Up to now, quite a few papers have been devoted to the computation of the supremal sublanguages that occur in SCT, either for specific properties with guaranteed finite convergence~\cite{WC18},~\cite{BGRLMW90},~\cite{W87},~\cite{CZW2014},~\cite{ZMW05},~\cite{ZC94},~\cite{CL91},~\cite{CM89},~\cite{KS05}, or in a systematic approach~\cite{ZKW99},~\cite{KG95a}, which sometimes has to  sidestep the finite  convergence issue in the  discussions. 




In this paper, we adopt the tool of  (semi)-topology for computing some supremal sublanguages which occur in the supervisory control theory. The basic idea is to identify the solutions of the defining language equations as open sets for some (semi)-topologies. Then, the supremal sublanguages naturally correspond to the supremal open subsets, i.e., the interiors. Essentially, several properties of interest, such as controllability and normality, are open subsets for different topologies. This naturally leads to the perspective that “supervisors are  open subsets". We hope this work may help to point out the  usefulness of elementary topology, which has not found serious applications in  the supervisory control theory, for the synthesis of supervisors. 

This paper is to be organized as follows. Section~\ref{sec: MP} provides some preliminaries on the supervisory control theory and others.   In Section~\ref{se: ST}, we introduce the basic tool of (semi)-topology to support the computation of some supremal sublanguages. In Section~\ref{sec: examples}, we provide examples to illustrate the topological techniques.  We  conclude the paper in Section~\ref{sec: Con}. 

\section{Mathematical Preliminaries}
\label{sec: MP}
In this section, we recall most of the notations and terminologies that will be used in this paper, from the supervisory control theory and others.

Let $\Sigma$ denote an alphabet. For any two languages $L_1, L_2$ over $\Sigma$, their concatenation is denoted by $L_1L_2$.
We write $L_1\backslash L_2:=\{s \in L_1 \mid s \notin L_2\}$ to denote their set-theoretic difference. The right quotient of $L_1$ by $L_2$ is denoted by $L_1/L_2:=\{s \in \Sigma^* \mid \exists t \in L_2, st \in L_1\}$, where $st$ denotes the string concatenation of $s$ and $t$. The prefix-closure  of $L$ is denoted by $\overline{L}$. The Kleene closure of $L$ is denoted by $L^*$.   We sometimes denote the complement of $L$ w.r.t. $\Sigma^*$ as $L^c$, i.e., $L^c:=\Sigma^* \backslash L$.  

Let $\Sigma_o \subseteq \Sigma$. We define natural projection $P_{\Sigma_o}: \Sigma^* \longrightarrow \Sigma_o^*$ recursively as follows.
\begin{enumerate}
    \item $P_{\Sigma_o}(\epsilon)=\epsilon$, where $\epsilon$ denotes the empty string,
    \item for any $s \in \Sigma^*$ and any $\sigma \in \Sigma$, $P_{\Sigma_o}(s\sigma)=P_{\Sigma_o}(s)$ if $\sigma \notin \Sigma_o$, and $P_{\Sigma_o}(s\sigma)=P_{\Sigma_o}(s)\sigma$ if $\sigma \in \Sigma_o$.
\end{enumerate}
The definition of $P_{\Sigma_o}$ is extended to  $P_{\Sigma_o}: 2^{\Sigma^*} \longrightarrow 2^{\Sigma_o^*}$ in such a way that $P_{\Sigma_o}(L):=\{P_{\Sigma_o}(s) \in \Sigma_o^* \mid s \in L\}$ for any $L \subseteq \Sigma^*$. The inverse projection $P_{\Sigma_o}^{-1}: \Sigma_o^* \longrightarrow 2^{\Sigma^*}$ is defined such that $P_{\Sigma_o}^{-1}(t):=\{s \in \Sigma^* \mid P_{\Sigma_o}(s)=t\}$ for any $t \in \Sigma_o^*$. Similarly, we define $P_{\Sigma_o}^{-1}(L')=\bigcup_{t \in L'}P_{\Sigma_o}^{-1}(t)$ for any $L' \subseteq \Sigma_o^*$, as a natural extension.

A binary relation $I \subseteq \Sigma \times \Sigma$
is said to be an independence relation (over $\Sigma$) if it is both irreflexive and symmetric.
$I$ is irreflexive if, for any  $\sigma \in \Sigma$, $(\sigma, \sigma) \notin I$. $I$ is  symmetric if, for any $\sigma, \sigma' \in \Sigma$, $(\sigma, \sigma') \in I$ if and only if $(\sigma', \sigma) \in I$.  For any
$(\sigma, \sigma') \in I$, $\sigma$ and $\sigma'$ are said to be independent symbols. 

Two strings $s, s'$ over $\Sigma$ are said to be trace equivalent with
respect to (the independence relation) $I$, denoted by $s \sim_{I} s'$,
if there exist strings $v_0, \ldots, v_k$ for some $k \geq 0$ such that $s=v_0$, $s'=v_k$ and
for each $i \in [1, k]$, there exist some $u_i, u_i' \in \Sigma^*$ and $\sigma_i, \beta_i \in \Sigma$
such that $(\sigma_i, \beta_i) \in I$, $v_{i-1}=u_i\sigma_i\beta_iu_i'$ and $v_i = u_i\beta_i\sigma_iu_i'$. Intuitively, two strings are trace equivalent if each one of them
can be obtained from the other by a sequence of permutations
of adjacent symbols that are independent. 

The set of trace
equivalent strings of $s$ for $I$ is said to be the trace closure of
$s$ with respect to $I$, denoted by $[s]_I$. The trace closure $[L]_I$
of a language $L \subseteq  \Sigma^*$
is defined to be the language $\bigcup_{s \in L}[s]_I$.

We shall remark that the operations of taking complement, intersection,  prefix-closure, projection and inverse projection preserve regularity of languages. However, the trace closure operation does not preserve the regularity of languages~\cite{AW86}. 

Several important language-theoretic properties in SCT are formulated as language equations given below.

\begin{definition} [Normality] Let $K \subseteq L \subseteq \Sigma^*$  and  $\Sigma_o\subseteq \Sigma$. $K$ is normal w.r.t.
$(L, \Sigma_o)$ if $K = P_{\Sigma_o}^{-1}P_{\Sigma_o}(K) \cap L$. 
\end{definition}
\begin{definition}[Controllability] Let $K \subseteq L=\overline{L} \subseteq \Sigma^*$  and 
$\Sigma_{uc}\subseteq \Sigma$. $K$ is said to be controllable
w.r.t. $(L, \Sigma_{uc})$ if $\overline{K}\Sigma_{uc}^* \cap L=\overline{K}$.
\end{definition}
\begin{definition} [$L$-closedness] Let $K \subseteq L \subseteq \Sigma^*$. $K$ is said to be
$L$-closed if $K = \overline{K} \cap L$.
\end{definition}
\begin{definition}
[Trace-closedness] Let $K \subseteq \Sigma^*$. $K$ is said to be trace-closed w.r.t. an independence relation $I$ over $\Sigma$ if $K=[K]_I$
\end{definition}
\begin{definition}
[Prefix-closedness] Let $K \subseteq \Sigma^*$. $K$ is said to be prefix-closed if $K=\overline{K}$.
\end{definition}

The following straightforward result holds.
\begin{lemma}
\label{lema: prefix}
 $K$ is prefix-closed iff $K$ is $\Sigma^*$-closed.
\end{lemma}
\section{Semi-Topologies for Computation}
\label{se: ST}

Let $M \subseteq \Sigma^*$ be any language. A function $\square: 2^M \longrightarrow 2^M$
is said to be a semi-topological closure
operator (on 
$M$) if it satisfies the following four  axioms S1, S2, S3, S4: for any $A, B \subseteq M$, 
\begin{enumerate}
\item[(S1):] $A \subseteq A^{\square}$,
\item[(S2):] $(A^{\square})^{\square}=A^{\square}$,
\item[(S3):] $A^{\square} \cup B^{\square} \subseteq (A \cup B)^{\square}$,
\item[(S4):] $\varnothing^{\square}=\varnothing$,
\end{enumerate}
where S1, S2, S3 are adopted from~\cite{XX}. We here remark that S3 here only requires $A^{\square} \cup B^{\square} \subseteq (A \cup B)^{\square}$ to hold, compared with the stronger requirement that $A^{\square} \cup B^{\square} =(A \cup B)^{\square}$ in the Kuratowski closure axioms that define a topological closure operator~\cite{James}.  

The semi-topological closure operator $\square$ on $M$
induces a
(unique) semi-topology $(M, \mathcal{T})$ with $\mathcal{T} = \{M\backslash L^{\square} \mid L \subseteq M\}$ being  the family of open sets such
that the following three conditions T1, T2, T3 hold.
\begin{enumerate}
\item [(T1):] $\varnothing \in \mathcal{T}$,
\item [(T2):] $\mathcal{T}$ is closed under arbitrary unions,
\item [(T3):] $M \in \mathcal{T}$.
\end{enumerate}
On the other hand, $L \subseteq M$  is said to be closed  if $L =
L^{\square}$. $\{L^{\square} \mid L \subseteq M\}$ is the  family of closed sets. In particular, $K \subseteq M$ is open (respectively, closed) iff $M\backslash K$ is closed (respectively, open), in the semi-topology $(M, \mathcal{T})$.

The semi-topological interior operator $\circ: 2^M \longrightarrow 2^M$ can be defined and is dual to the  semi-topological closure operator $\square$, with $L^{\circ}=M \backslash (M\backslash L)^{\square}$ for any $L \subseteq M$. It satisfies the following four conditions S1', S2', S3', S4': for any $A, B \subseteq M$,
\begin{enumerate}
\item[(S1'):] $A \supseteq A^{\circ}$,
\item[(S2'):] $(A^{\circ})^{\circ}=A^{\circ}$,
\item[(S3'):] $A^{\circ} \cap B^{\circ} \supseteq (A \cap B)^{\circ}$,
\item[(S4'):] $M^{\circ}=M$,
\end{enumerate}
The conditions S1', S2', S3', S4' are, respectively, dual to and can be obtained from the axioms 
S1, S2, S3, S4.

In the following subsections, we consider four increasingly more powerful  techniques, however, with the same basic topological idea.
\subsection{Supremal Solution of A Topologized Equation}
\label{sec: atopo}

We have the next useful result, which is a straightforward consequence of the semi-topology  endowed on $M$.

\begin{theorem}
\label{them: supremal}
There exists the unique supremal open subset of $E$ for any language $E \subseteq M$. Indeed, the supremal open subset of $E$ is the interior $E^{\circ}=M\backslash (M\backslash E)^{\square }$ of $E$.
\end{theorem}

\emph{Remark}:
To make the paper more self-contained, we here briefly explain why the interior $M\backslash (M\backslash E)^{\square }$ is indeed the supremal open subset of $E$ in the semi-topology $(M, \mathcal{T})$. Let $M\backslash L^{\square}  \subseteq E$ be any given open subset of $E$. We have $L^{\square} \supseteq M \backslash E$, from $M\backslash L^{\square  }\subseteq E$,   and thus $L^{\square}\supseteq (M\backslash E)^{ \square}$. It then follows that $M\backslash L^{\square }\subseteq M \backslash (M\backslash E)^{ \square}$. $M \backslash (M\backslash E)^{ \square}$ is clearly an open subset of $E$. 

Indeed, in the terminology of the rough set theory~\cite{P92}, $M\backslash (M\backslash E)^{\square }$ is the largest under-approximation of $E$ in the collection of open subsets of $E$. 


The next result could be more convenient in some applications, which is a direct consequence of Theorem~\ref{them: supremal}.
\begin{corollary}
\label{coral: closed}
    If the family of open
sets corresponds to the family of closed sets for
the semi-topology induced by $\square$ on $M$, then the supremal closed subset of $E$ is $M\backslash (M\backslash E)^{\square }$. 
\end{corollary}


Corollary~\ref{coral: closed} states that, if the semi-topology induced by $\square$ on $M$ satisfies the following property  ($*$): the family of open  sets corresponds  to the family of closed sets, then $M\backslash (M\backslash E)^{\square }$ is also the largest under-approximation of $E$ in the collection of closed subsets of $E$. We shall refer to any (semi-topological) closure operator that satisfies the  property ($*$) as a clopen closure operator.

 Theorem~\ref{them: supremal} can be used in the following way. Suppose we can formulate a property as the solutions of an equation $K=M\backslash (M\backslash K)^{\square}$, with $K \subseteq E \subseteq M$ and a function $\square: 2^M \longrightarrow 2^M$. If $\square$ is a semi-topological closure operator, then this equation states that $K$ is equal to its interior $K^{\circ}$, that is, $K$ is open, in the semi-topology induced by $\square$ on $M$. Then, the supremal solution of this equation exists and is equal to the interior $E^{\circ}=M\backslash (M \backslash E)^{\square}$ of $E$. 

Corollary~\ref{coral: closed} can be used in a similar manner. Suppose we can formulate a property as the solutions of an equation $K= K^{\square}$, with $K \subseteq E \subseteq M$ and a function $\square: 2^M \longrightarrow 2^M$. If $\square$ is a  clopen closure operator, then the supremal solution of this equation exists and is equal to the interior $E^{\circ}=M\backslash (M \backslash E)^{\square}$ of $E$.

\emph{Remark}:
If $\square$ is a clopen closure operator, then $K=K^{\square}$ ($K$ is a closed set) if and only if $K=M\backslash (M\backslash K)^{\square}=K^{\circ}$ ($K$ is an open set). 

In the rest of this paper, for generality, we mostly only consider equations of the form $K=M\backslash (M\backslash K)^{\square}$. However, we shall work with the equation  $K=K^{\square}$ if $\square$ is a clopen closure operator and 
it is more convenient to do so.
\subsection{Supremal Solution of  System of Topologized Equations}
\label{sec: system}
Let $M_i \subseteq \Sigma^*$ be any given language and let $\square_i: 2^{M_i} \longrightarrow 2^{M_i}$ be any semi-topological closure operator on $M_i$, for each $i \in [1, n]$. Let $E \subseteq \bigcap_{i=1}^nM_i$.

Consider a system $\Delta$ of language equations in the topologized forms
\begin{align*}
  K=M_1\backslash (M_1\backslash K)^{ \square_1 }, \\ 
  K=M_2\backslash (M_2\backslash K)^{ \square_2 }, \\ 
  \ldots \\
  K=M_n\backslash (M_n\backslash K)^{ \square_n },
\end{align*}
with $K \subseteq E$. Thus, any solution of this system is an open set
in the semi-topology induced by  $\square_i$ on $M_i$, for each $i \in [1, n]$.
 By T2, the supremal solution exists, and it can be successively approximated with the iteration scheme
$$K_{i+1}=K_i^{\circ_1 \circ_2\ldots \circ_n},$$
where $K_0=E$, and each function $\circ_i: 2^{M_i} \longrightarrow 2^{M_i}$ is a semi-topological interior operator that is dual to $\square_i$,
with $K^{\circ_i}:=M_i\backslash (M_i\backslash K)^{\square_i}$ for any $K \subseteq M_i$. The correctness of the iteration scheme is shown in the following.
\begin{theorem}
Let $K^{\uparrow}$ denote the supremal solution of the system $\Delta$ of equations. If the iteration scheme
\begin{center}$K_{i+1}=K_i^{\circ_1 \circ_2\ldots \circ_n}$, with
$K_0=E$
\end{center}terminates with $K_l$, where $l\geq 0$, then $K_l=K^{\uparrow}$.
\end{theorem}
\emph{Proof}: The descending chain $$E=K_0 \supseteq K_1 \supseteq \ldots \supseteq K_l =K_{l+1}=\ldots$$ is obtained. We first show that  $K^{\uparrow } \subseteq K_i$ for each $i \geq 0$.

By definition, $(K^{\uparrow})^{\circ_j}=K^{\uparrow}$ for each $j \in [1, n]$. Thus, for each $i \geq 0$, $K^{\uparrow} \subseteq K_i$ implies
$K^{\uparrow} \subseteq K_i^{\circ_1 \circ_2 \ldots \circ_n}=K_{i+1}$. The claim $K^{\uparrow } \subseteq K_i$, for each $i \geq 0$, follows from $K^{\uparrow } \subseteq E=K_0$. Thus, we have $K^{\uparrow}\subseteq K_l$.

Next, we show  $K^{\uparrow}\supseteq K_l$. By definition, $K_l=K_{l+1}=K_l^{\circ_1 \circ_2 \ldots \circ_n}$. Thus, we have $K_l^{\circ_j} \subseteq K_l=K_l^{\circ_1 \circ_2 \ldots \circ_n} \subseteq K_l^{\circ_j}$, for each $j \in [1, n]$. That is, $K_l=K_l^{\circ_j}$ for each $j \in [1, n]$. With $K_l \subseteq E$, we know that $K_l$ is also a solution of the system $\Delta$ of equations. Thus, $K^{\uparrow}\supseteq K_l$.
\hfill $\blacksquare$


\subsection{Supremal Solution of  System of Topologized Equations with Search Space Relaxation}
\label{sec: relax}

In some cases, the  topological approach given above cannot be directly applied. This is so, for example, when we need to search for the supremal solution of $K \subseteq E$, while the topological characterization only applies to $\overline{K}$. For example, consider the following system $\Theta$ of equations \begin{align*}
  \overline{K}=M_1\backslash (M_1\backslash \overline{K})^{ \square_1 }, \\ 
  \overline{K}=M_2\backslash (M_2\backslash \overline{K})^{ \square_2 }, \\ 
  \ldots \\
  \overline{K}=M_n\backslash (M_n\backslash \overline{K})^{ \square_n },
\end{align*}
where $K \subseteq E$. As before, $M_i \subseteq \Sigma^*$ is any given language and  $\square_i: 2^{M_i} \longrightarrow 2^{M_i}$ is a semi-topological closure operator (on $M_i$), for each $i \in [1, n]$. Here, $\overline{E} \subseteq \bigcap_{i=1}^nM_i$. The supremal solution for this system exists (see Proposition~\ref{prop: exists} for a generalization).

In this case, the use of the weakened requirement $\overline{K} \subseteq \overline{E}$ (for the search space) allows one to apply the topological techniques developed before for $\overline{K}$.
However, this relaxation only computes the supremal solution of $\overline{K}$ from the topologized equations. The solution  needs to be constrained as we only ask for the supremal solution of $K$, for which it holds that $K\subseteq \overline{K} \cap E$. This leads to a natural iteration scheme that successively approximates the supremal solution of $K$, subjected to the language equations for $\overline{K}$. Now, consider a more general setup, where we have a system $\Lambda$ of  equations 

\begin{align*}
  K^{\square}=M_1\backslash (M_1\backslash K^{\square})^{ \square_1 }, \\ 
  K^{\square}=M_2\backslash (M_2\backslash K^{\square})^{ \square_2 }, \\ 
  \ldots \\
  K^{\square}=M_n\backslash (M_n\backslash K^{\square})^{ \square_n },
\end{align*}
where $K \subseteq E$. Here, $M_i$ and $\square_i$ are the same as before. 

We assume $\square: 2^M \longrightarrow 2^M$ is a semi-topological closure operator on some $M\subseteq \Sigma^*$ such that the equation $X=X^{\square}$ has an equivalent dual $X=M\backslash (M\backslash X)^{ \square '}$ for some semi-topological closure operator $\square': 2^M \longrightarrow 2^M$.
 Let $E^{\square} \subseteq \bigcap_{i=1}^nM_i$. 

Then, the following iteration scheme ($\#$) works for the system $\Lambda$, by replacing $K^{\square}$ with $X$.
\begin{enumerate}
\item Let $i=0$ and $K_i=E$.

\item Solve the following system $\Gamma(i)$ of equations 
\begin{align*}
X=M\backslash (M\backslash X)^{ \square '}, \\
X=M_1\backslash (M_1\backslash X)^{ \square_1 },\\
X=M_2\backslash (M_2\backslash X)^{ \square_2 },\\
\ldots,\\
X=M_n\backslash (M_n\backslash X)^{ \square_n },
\end{align*}
where $X \subseteq K_{i}^{\square}$. The supremal solution  $X_i$ exists, as  for the system $\Delta$, and can be computed with the  iteration scheme as given below.
$$L_{j+1}=L_j^{\circ_1 \circ_2\ldots \circ_n \circ'},$$ where $L_0=K_i^{\square}$. Let $K_{i+1}=K_i \cap X_i$. Let $i:=i+1$ and repeat step 2).
\end{enumerate}
\emph{Remark}:
The system $\Lambda$ contains the system $\Theta$ as a special case. Indeed,  let $\square: 2^{\Sigma^*} \longrightarrow 2^{\Sigma^*}$ be defined such that $K^{\square}=\overline{K}$ for any $K \subseteq \Sigma^*$. $\square$ is a semi-topological closure operator on $\Sigma^*$ and the equation $X=X^{\square}$ has an equivalent dual $X=\Sigma^*\backslash (\Sigma^* \backslash X)^{\square'}$, where $\square':2^{\Sigma^*} \longrightarrow 2^{\Sigma^*}$ is a semi-topological closure operator on $\Sigma^*$ defined such that $L^{\square'}=L\Sigma^*$ for any $L \subseteq \Sigma^*$.  Indeed, we have $X=\overline{X}$ iff $\overline{X}\subseteq X$ iff $\overline{X} \cap (\Sigma^*\backslash X)=\varnothing$ iff $X \cap (\Sigma^*\backslash X)\Sigma^*=\varnothing$ iff $X \subseteq \Sigma^*\backslash (\Sigma^* \backslash X)\Sigma^*$ iff $X = \Sigma^*\backslash (\Sigma^* \backslash X)\Sigma^*$. The system $\Lambda$ also contains the system $\Delta$ as a special case, by setting $K^{\square}=K$ for any $K \subseteq \Sigma^*$.

Before we show the correctness of the  iteration scheme given above, we first show some useful results.
\begin{proposition}
Suppose $\square:2^M \longrightarrow 2^M$ is a clopen closure operator on $M$, then  $X=X^{\square}$ has an equivalent dual $X=M\backslash (M\backslash X)^{ \square '}$ for some  semi-topological closure operator $\square': 2^M \longrightarrow 2^M$.
\end{proposition}
\emph{Proof}: Suppose $\square: 2^M \longrightarrow 2^M$ is a clopen closure operator on $M$. Then, for any language $L \subseteq M$,  $L$ satisfies the equation  $X=X^{\square}$ (that is, $L$ is a closed set) iff $L$ satisfies the equation $X=M\backslash (M\backslash X)^{ \square}$ (that is, $L$ is an open set). We can choose $\square':=\square$. \hfill $\blacksquare$ 

\emph{Remark}:
 $\square$ being a clopen closure operator on $M$ is a sufficient, but in general not  necessary, condition for $X=X^{\square}$ to have an equivalent dual $X=M\backslash (M\backslash X)^{ \square '}$, for some semi-topological closure operator $\square'$. For example, $K=\overline{K}$ has an equivalent dual $K=\Sigma^*\backslash (\Sigma^* \backslash K)\Sigma^*$ but the prefix closure operator is a semi-topological closure operator on $\Sigma^*$ that is not clopen.

We have the next useful result.
\begin{lemma}
\label{lemma:union}
Suppose the equation $X=X^{\square}$ has an equivalent dual $X=M\backslash (M\backslash X)^{ \square '}$, where $\square$ and $\square'$ are semi-topological closure operators on $M$, then $(\bigcup_{\alpha \in I}K_{\alpha})^{\square}=\bigcup_{\alpha \in I}K_{\alpha}$ holds for any index set $I$,  if each $K_{\alpha}$ is a closed set, i.e., $K_{\alpha}=K_{\alpha}^{\square}$. 
\end{lemma}
\emph{Proof}:
Each $K_{\alpha}$ satisfies the equation $X=X^{\square}$. Thus, $K_{\alpha}=M\backslash (M\backslash K_{\alpha})^{ \square '}$ for each $\alpha \in I$. That is, each $K_{\alpha}$ is an open set for the semi-topology on $M$ induced by $\square'$. It follows that $\bigcup_{\alpha \in I}K_{\alpha}$ is also an open set and thus it satisfies the equation $X=M\backslash (M\backslash X)^{ \square '}$. It then follows that $\bigcup_{\alpha \in I}K_{\alpha}$ satisfies the equation $X=X^{\square}$, that is, $\bigcup_{\alpha \in I}K_{\alpha}$ is a closed set in the semi-topology induced by $\square$. The statement holds even when $I$ is an empty index set, for which we have $\bigcup_{\alpha \in I}K_{\alpha}=\varnothing$.
\hfill $\blacksquare$\\ 

We now show the following result.
\begin{proposition}
\label{prop: exists}
The supremal solution $K^{\uparrow}$ for the system $\Lambda$ exists.
\end{proposition}
\emph{Proof}:
Since $\square$ is a semi-topological closure operator on $M$, we have $\varnothing^{\square}=\varnothing$; we have that $\varnothing$ is a solution of the system $\Lambda$. Let $\{K_{\alpha}\mid \alpha \in I\}$ be any (non-empty) set of solutions of the system $\Lambda$. We show that $\bigcup \{K_{\alpha}\mid \alpha \in I\}$ is also a solution of the system $\Lambda$. We only need to show that, for each $i \in [1, n]$, $$ (\bigcup \{K_{\alpha}\mid \alpha \in I\})^{\square} \subseteq M_i\backslash (M_i \backslash (\bigcup \{K_{\alpha}\mid \alpha \in I\})^{\square})^{\square_i}. $$
We have that $(\bigcup \{K_{\alpha}\mid \alpha \in I\})^{\square}\supseteq \bigcup_{\alpha \in I} K_{\alpha}^{\square}$.  Thus, $$M_i \backslash (\bigcup \{K_{\alpha}\mid \alpha \in I\})^{\square} \subseteq \bigcap _{\alpha \in I} (M_i\backslash K_{\alpha}^{\square}).$$
It follows that 
\begin{center}$(M_i \backslash (\bigcup \{K_{\alpha}\mid \alpha \in I\})^{\square})^{\square_i} \subseteq (\bigcap _{\alpha \in I} (M_i\backslash K_{\alpha}^{\square}))^{\square_i} \subseteq \bigcap_{\alpha \in I} (M_i\backslash K_{\alpha}^{\square})^{\square_i}.$\end{center}
Thus, 
\begin{center}
    $M_i\backslash (M_i \backslash (\bigcup \{K_{\alpha}\mid \alpha \in I\})^{\square})^{\square_i}\supseteq M_i \backslash \bigcap_{\alpha \in I} (M_i\backslash K_{\alpha}^{\square})^{\square_i}=\bigcup_{\alpha \in I}M_i \backslash (M_i\backslash K_{\alpha}^{\square})^{\square_i}=\bigcup_{\alpha \in I}K_{\alpha}^{\square}$.
\end{center}
By Lemma~\ref{lemma:union}, we have that $$\bigcup_{\alpha \in I}K_{\alpha}^{\square}=(\bigcup_{\alpha \in I}K_{\alpha}^{\square})^{\square} \supseteq (\bigcup \{K_{\alpha}\mid \alpha \in I\})^{\square}.$$
This finishes the proof.\hfill $\blacksquare$\\

The correctness of the  successive approximation scheme is shown below.

\begin{theorem}
Let $K^{\uparrow}$ denote the supremal solution of the system $\Lambda$ of equations. If the iteration scheme ($\#$)
 terminates with $K_l$, where $l\geq 0$, then  $K_l=K^{\uparrow}$.
\end{theorem}
\emph{Proof}:
The descending chain $$E=K_0 \supseteq K_1  \supseteq \ldots  \supseteq K_l=K_{l+1}=\ldots$$ is obtained from the iteration scheme. We first show that  $K_i \supseteq K^{\uparrow}$ for each $i \geq 0$.

To that end, we shall show that, for any $i \geq 0$, $K_i \supseteq K^{\uparrow}$ implies $K_{i+1} \supseteq K^{\uparrow}$. Since $K_{i+1}=K_i \cap X_i$, we only need to show that  $K_i \supseteq K^{\uparrow}$ implies $X_i \supseteq K^{\uparrow}$, where $X_i$ is the supremal solution of the system $\Gamma(i)$ of equations
\begin{align*}
X=M\backslash (M\backslash X)^{ \square '}, \\
X=M_1\backslash (M_1\backslash X)^{ \square_1 },\\
X=M_2\backslash (M_2\backslash X)^{ \square_2 },\\
\ldots,\\
X=M_n\backslash (M_n\backslash X)^{ \square_n },
\end{align*}
where $X \subseteq K_{i}^{\square}$.

Suppose $K^{\uparrow} \subseteq K_i$, we have $(K^{\uparrow})^{\square} \subseteq K_i^{\square}$. Since $K^{\uparrow}$ is the supremal solution of the system $\Lambda$ and  $(K^{\uparrow})^{\square} \subseteq K_i^{\square}$, we conclude that $(K^{\uparrow})^{\square}$ is a solution of the system $\Gamma(i)$. Thus, $(K^{\uparrow})^{\square} \subseteq X_i$. Consequently, we have $K^{\uparrow} \subseteq X_i$.

This finishes the proof of the claim that, for any $i \geq 0$, $K_i \supseteq K^{\uparrow}$ implies $K_{i+1} \supseteq K^{\uparrow}$. Now, since $K^{\uparrow} \subseteq E=K_0$, we conclude that $K_i \supseteq K^{\uparrow}$ for each $i \geq 0$.

We have $K_l=K_{l+1}=K_{l} \cap X_{l}$. Thus, we have $K_{l} \subseteq X_{l}$. We recall that $X_{l}$ is the supremal solution of the system $\Gamma(l)$ of equations
\begin{align*}
X=M\backslash (M\backslash X)^{ \square '}, \\
X=M_1\backslash (M_1\backslash X)^{ \square_1 },\\
X=M_2\backslash (M_2\backslash X)^{ \square_2 },\\
\ldots,\\
X=M_n\backslash (M_n\backslash X)^{ \square_n },
\end{align*}
where $X \subseteq K_{l}^{\square}$, which is computed with the iteration $$L_{j+1}=L_j^{\circ_1 \circ_2\ldots \circ_n \circ'},$$ with $L_0=K_{l}^{\square}$. Since the computation terminates, there is  some $m \geq 0$ such that $X_{l}=K_{l}^{\square (\circ_1 \circ_2\ldots \circ_n \circ')^m}$. Thus,
$$K_l \subseteq X_l= K_{l}^{\square (\circ_1 \circ_2\ldots \circ_n \circ')^m}\subseteq K_{l}^{\square \circ_1 \circ_2\ldots \circ_n \circ'}$$
Let $T:=K_{l}^{\square \circ_1 \circ_2\ldots \circ_n\circ'}$. We have $T=T^{\circ'}=M\backslash (M\backslash T)^{ \square' }$. Since  $X=X^{\circ'}=M\backslash (M\backslash X)^{ \square' }$ has an equivalent dual $X=X^{\square}$, we have  $T=T^{\square}$. Thus, from $K_l \subseteq T$, we obtain $K_l^{\square} \subseteq T^{\square}=T=K_{l}^{\square \circ_1 \circ_2\ldots \circ_n\circ'}$. We then conclude that
\begin{align*}
K_l^{\square}\subseteq K_l^{\square \circ_1} \subseteq K_l^{\square}, \\
K_l^{\square}\subseteq K_l^{\square \circ_2} \subseteq K_l^{\square},\\
\ldots,\\
K_l^{\square}\subseteq K_l^{\square \circ_n} \subseteq K_l^{\square},
\end{align*}
where $K_l \subseteq E$. Thus, $K_l$ is a solution of the system $\Lambda$ and we have $K_l \subseteq K^{\uparrow}$. This, when combined with the fact that $ K^{\uparrow}  \subseteq K_l$, shows that $ K_l=K^{\uparrow}$. This finishes the proof.\hfill $\blacksquare$\\

\subsection{Supremal Solution of Mixed System of  Topologized Equations}
\label{sec: mixed}
In general, consider the system $\Xi$ of equations
\begin{align*}
  K^{\square_1'}=M_1\backslash (M_1\backslash K^{\square_1'})^{ \square_1 }, \\ 
  K^{\square_2'}=M_2\backslash (M_2\backslash K^{\square_2'})^{ \square_2 }, \\ 
  \ldots \\
  K^{\square_n'}=M_n\backslash (M_n\backslash K^{\square_n'})^{ \square_n },
\end{align*}
with $K \subseteq E$. Let $M_i \subseteq \Sigma^*$ be any given language and  $\square_i: 2^{M_i} \longrightarrow 2^{M_i}$ be a semi-topological closure operator (on $M_i$), for each $i \in [1, n]$. For each $i \in [1, n]$, let $\square_i'$ be a semi-topological closure operator on some $M_i' \subseteq \Sigma^*$ such that the equation $X=X^{\square_i'}$ has an equivalent dual $X=M_i' \backslash (M_i'\backslash X)^{\square_i''}$ for some semi-topological closure operator $\square_i''$ on $M_i'$. Here, $E^{\square_i'} \subseteq M_i$  for each $i \in [1, n]$.

\emph{Remark}: The system $\Xi$ is a straightforward generalization of the system $\Lambda$. In particular, if $\square_1'=\square_2'=\ldots=\square_n'$, then the system $\Xi$ degenerates into the system $\Lambda$.

By an analysis which is similar to the proof of Proposition~\ref{prop: exists}, the supremal solution $K^{\uparrow}$ of the system $\Xi$ exists. A strategy to solve the system $\Xi$ is to iteratively solve each equation, by using the technique developed in Section~\ref{sec: relax}. This leads to a natural iteration scheme shown below.

Given any equation $K^{\square_i'}=M_i\backslash (M_i\backslash K^{\square_i'})^{ \square_i },$ with $K \subseteq L$. The supremal solution exists and is denoted by $L^{\diamond_i}$ for convenience, which can be computed with the technique developed in Section~\ref{sec: relax}. Then, the system $\Xi$ is solved with the following iteration scheme 
\begin{center}
    $K_{j+1}=K_j^{\diamond_1\diamond_2\ldots \diamond_n}$,
\end{center}
with $K_0=E$. The correctness of the iteration scheme is shown below.
\begin{theorem}
Let $K^{\uparrow}$ denote the supremal solution of the system $\Xi$ of equations. If the iteration scheme \begin{center}
    $K_{j+1}=K_j^{\diamond_1\diamond_2\ldots \diamond_n}$, with $K_0=E$
\end{center}
terminates with $K_l$, where $l \geq 0$, then $K_l=K^{\uparrow}$.
\end{theorem}
\emph{Proof}: The descending chain $$E=K_0 \supseteq K_1  \supseteq \ldots  \supseteq K_l=K_{l+1}=\ldots$$ is obtained from the iteration scheme. We first show that  $K_i \supseteq K^{\uparrow}$ for each $i \geq 0$.

To that end, we shall show that, for any $i \geq 0$, $K_i \supseteq K^{\uparrow}$ implies $K_{i+1} \supseteq K^{\uparrow}$. Indeed, suppose $K_i \supseteq K^{\uparrow}$. Then, $K^{\uparrow}$ is a solution of the system of equations
\begin{align*}
  K^{\square_1'}=M_1\backslash (M_1\backslash K^{\square_1'})^{ \square_1 }, \\ 
  K^{\square_2'}=M_2\backslash (M_2\backslash K^{\square_2'})^{ \square_2 }, \\ 
  \ldots \\
  K^{\square_n'}=M_n\backslash (M_n\backslash K^{\square_n'})^{ \square_n },
\end{align*}
where $K \subseteq K_i$. In particular, we have $K^{\uparrow}$ is a solution of the equation $K^{\square_1'}=M_1\backslash (M_1\backslash K^{\square_1'})^{ \square_1 },$ with $K \subseteq K_i$. Thus, $K^{\uparrow} \subseteq K_i^{\diamond_1}$. It follows that $K^{\uparrow}$ is a solution of the equation $K^{\square_2'}=M_2\backslash (M_2\backslash K^{\square_2'})^{ \square_2 },$ with $K \subseteq K_i^{\diamond_1}$. We then have $K^{\uparrow} \subseteq K_i^{\diamond_1\diamond_2}$. The same reasoning shows that $K^{\uparrow} \subseteq K_i^{\diamond_1\diamond_2\ldots \diamond_n}=K_{i+1}$.

The claim  that $K_i \supseteq K^{\uparrow}$ for each $i \geq 0$ then follows from $K^{\uparrow} \subseteq E=K_0$. Thus, $K_l \supseteq K^{\uparrow}$.

We have $K_l=K_{l+1}=K_l^{\diamond_1\diamond_2\ldots \diamond_n} \subseteq K_l^{\diamond_i} \subseteq K_l$, for each $i \in [1, n]$. Thus, we have $K_l=K_l^{\diamond_i}$, for each  $i \in [1, n]$. Thus, $K_l$ is a solution of the system $\Xi$ and $K_l \subseteq K^{\uparrow}$.
\hfill $\blacksquare$\\

\section{Illustrative Examples}
\label{sec: examples}
In this section, we shall provide some examples to illustrate the previously presented ideas.
\subsection{Computation With A Topologized Equation}
In this subsection, we present some example properties which can be specified with a topologized equation, including the properties of normality, $L$-closedness, prefix-closedness and trace-closedness. It thus follows that the supremal sublanguages for these properties can be expressed with the same formula given in Section~\ref{sec: atopo}.
\subsubsection{Supremal Normal Sublanguage}
Let $$N(E;L_m(G)):=\{K \subseteq E \mid K = P_{\Sigma_o}^{-1}P_{\Sigma_o}(K) \cap L_m(G)\}$$ be the collection of normal sublanguages of $E$ with
respect to $(L_m(G), \Sigma_o)$. Without loss of generality,
we may assume that $E \subseteq L_m(G)$, since any $K \in N(E;L_m(G))$ satisfies the property that $K \subseteq L_m(G)$. We shall endow a (semi)-topology on $L_m(G)$. Let $\square_{N} : 2^{L_m(G)} \longrightarrow 2^{L_m(G)}$ denote a function, where, for any $K \subseteq L_m(G)$, $K^{\square_{N}}:=P_{\Sigma_o}^{-1}P_{\Sigma_o}(K) \cap L_m(G)$. $\square_{N}$ is a (semi)-topological closure operator on $L_m(G)$ (and is a clopen closure operator).

We reformulate the elements of $\mathcal{N}(E; L_m(G))$ as follows: with $K \subseteq E$, 
\begin{center}
$K = P_{\Sigma_o}^{-1}P_{\Sigma_o}(K) \cap L_m(G)$ iff $ P_{\Sigma_o}^{-1}P_{\Sigma_o}(K) \cap L_m(G) \subseteq K$ iff $P_{\Sigma_o}^{-1}P_{\Sigma_o}(K) \cap (L_m(G)\backslash K) =\varnothing$ iff $K \cap P_{\Sigma_o}^{-1}P_{\Sigma_o}(L_m(G)\backslash K)=\varnothing$ iff $K \subseteq L_m(G)\backslash P_{\Sigma_o}^{-1}P_{\Sigma_o}(L_m(G)\backslash K)$ iff $K= L_m(G)\backslash P_{\Sigma_o}^{-1}P_{\Sigma_o}(L_m(G)\backslash K)$ iff $K=L_m(G) \backslash (L_m(G)\backslash K)^{\square_N}$
\end{center}

Thus, ${\bf Sup}\mathcal{N}(E;L_m(G))$ exists in $\mathcal{N}(E;L_m(G))$ and 
\begin{center}${\bf Sup}\mathcal{N}(E;L_m(G))=L_m(G) \backslash (L_m(G)\backslash E)^{\square_N}=L_m(G)\backslash P_{\Sigma_o}^{-1}P_{\Sigma_o}(L_m(G) \backslash E).$\end{center}

\emph{Remark}:
It is straightforward to  show that this expression
is equivalent to the Lin-Brandt formula~\cite{WC18}
given by $${\bf Sup}\mathcal{N}(E;L_m(G))= E\backslash P_{\Sigma_o}^{-1}P_{\Sigma_o}(L_m(G) \backslash E)$$ (see also~\cite{BGRLMW90}). Indeed,
\begin{center}$L_m(G)\backslash P_{\Sigma_o}^{-1}P_{\Sigma_o}(L_m(G) \backslash E)=(E \cup (L_m(G)\backslash E)) \backslash P_{\Sigma_o}^{-1}P_{\Sigma_o}(L_m(G) \backslash E)=E\backslash P_{\Sigma_o}^{-1}P_{\Sigma_o}(L_m(G) \backslash E).$
\end{center}



\subsubsection {Supremal $L$-closed Sublanguage}

Let $$\mathcal{M}(E;L_m(G)):= \{K \subseteq E \mid K = \overline{K} \cap L_m(G)\}$$
denote the collection of $L_m(G)$-closed sublanguages of
$E$. Without loss of generality, we  assume 
$E \subseteq L_m(G)$. We now endow a
different topology on $L_m(G)$. Let $\square_L : 2^{L_m(G)} \longrightarrow
2^{L_m(G)}$ denote a function defined such that $K^{\square_L} := K\Sigma^* \cap L_m(G)$,  for
any $K \subseteq L_m(G)$.  $\square_L$ is a (semi)-topological
closure operator on $L_m(G)$. 

We shall reformulate
the elements of $\mathcal{M}(E;L_m(G))$ in the following manner:
with $K \subseteq E$, 
\begin{center}
$K = \overline{K} \cap L_m(G)$ iff $\overline{K} \cap L_m(G) \subseteq K$ iff $\overline{K} \cap (L_m(G)\backslash K)=\varnothing$
iff $K \cap (L_m(G)\backslash K)\Sigma^*=\varnothing$ 
iff $K \subseteq L_m(G)\backslash (L_m(G)\backslash K)\Sigma^*$ iff $K=  L_m(G)\backslash (L_m(G)\backslash K)\Sigma^*$ iff $K=  L_m(G)\backslash (L_m(G)\backslash K)^{\square_L}$
\end{center}
Thus, ${\bf Sup}\mathcal{M}(E;L_m(G))$ exists in $\mathcal{M}(E;L_m(G))$ and we have 
\begin{center}
${\bf Sup}\mathcal{M}(E;L_m(G)) =L_m(G)\backslash (L_m(G)\backslash E)^{\square_L}= L_m(G)\backslash (L_m(G)\backslash E)\Sigma^*.$
\end{center}

As a special case, we consider $L_m(G) = \Sigma^*$. Then, the supremal prefix-closed sublanguage ${\bf Sup}\mathcal{M}(E;\Sigma^*)$ of $E$ is $\Sigma^* \backslash (\Sigma^* \backslash E)\Sigma^*$, by Lemma~\ref{lema: prefix}.
In particular, the (semi)-topological closure operator for this case is denoted by $\square_P: 2^{\Sigma^*} \longrightarrow 2^{\Sigma^*}$, where $K^{\square_P}=K\Sigma^*$. 
\subsubsection{Supremal Trace-closed Sublanguage}
Let $I$ be an independence relation over $\Sigma$. Let
$$\mathcal{T}(E;  I):= \{K \subseteq E \mid K=[K]_I\}$$ denote the collection of trace-closed sublanguages of $E$. We now endow a topology on $\Sigma^*$. Let $\square_T: 2^{\Sigma^*} \longrightarrow 2^{\Sigma^*}$
be a map such that $K^{\square_T} := [K]_I$ for
any $K \subseteq \Sigma^*$.  $\square_T$ is a (semi)-topological
closure operator.
The complement of a trace-closed language w.r.t. $\Sigma^*$ is also trace-closed. Thus,  $\square_T$ is a clopen closure operator; ${\bf Sup}\mathcal{T}(E; I)$ exists in $\mathcal{T}(E; I)$  and 
\begin{center}
${\bf Sup}\mathcal{T}(E; I) =\Sigma^* \backslash (\Sigma^*\backslash E)^{\square_T}=\Sigma^* \backslash [\Sigma^*\backslash E]_I$.
\end{center}

However, the trace closure operator  does not always preserve the regularity of languages. Thus,  ${\bf Sup}\mathcal{T}(E; I)$ can be non-regular, even if $E$ is regular.


\subsection{Computation With A System of Topologized Equations}
\label{sec: to}
In this subsection, we use the property of prefix-closed controllability  to illustrate the computation with a system of topologized equations, as explained in Section~\ref{sec: system}, and explain an acceleration technique. 
\subsubsection{Supremal Prefix-closed Controllable Sublanguage}
\label{sec: pc}
Let $E=\overline{E}$ and let $$\overline{\mathcal{C}}(E;L(G)):=\{K \subseteq E\mid K=\overline{K}, K\Sigma_{uc}^*\cap L(G)=K\}.$$
denote the collection of prefix-closed controllable sublanguages of $E$ with respect to $(L(G), \Sigma_{uc})$. Without loss of generality, we assume $E \subseteq L(G)$. $\overline{\mathcal{C}}(E;L(G))$ is the solution set of a system of two language equations
\begin{align*}
K=\overline{K}, \\
K=K\Sigma_{uc}^*\cap L(G),
\end{align*}
with $K \subseteq E$.

We need to reformulate the
two language equations in the following manner:

1) $K=\overline{K}$ iff $K=\Sigma^* \backslash (\Sigma^*\backslash K)\Sigma^*$ iff $K=\Sigma^* \backslash (\Sigma^*\backslash K)^{\square_P}$

2) with $K \subseteq E$, $K\Sigma_{uc}^* \cap L(G) = K$ iff
$K\Sigma_{uc}^* \cap L(G) \subseteq K$ iff $K\Sigma_{uc}^* \cap (L(G)\backslash K) = \varnothing$
iff $K \cap (L(G)\backslash K)\slash \Sigma_{uc}^*=\varnothing$ iff $K \subseteq  L(G)\backslash ((L(G)\backslash K)\slash \Sigma_{uc}^*)$ iff $K=  L(G)\backslash ((L(G)\backslash K)\slash \Sigma_{uc}^*)$.

We shall endow a (semi)-topology on $L(G)$. We let $\square_C :
2^{L(G)} \longrightarrow 2^{L(G)}$ be a function defined such that $K^{\square_C} := K/\Sigma_{uc}^*
 \cap L(G)$,
for any $K \subseteq L(G)$.  $\square_C$ is indeed a (semi)-topological closure operator on $L(G)$.
Thus, $\overline{\mathcal{C}}(E;L(G))$ can be reformulated as the solution set of a system of two topologized  equations
\begin{align*}
K=\Sigma^*\backslash (\Sigma^*\backslash  K)^{ \square_P}, \\
K=L(G)\backslash (L(G)\backslash K)^{\square_C},
\end{align*}
with $K \subseteq E$. Then, the iteration scheme $$L_{i+1}=L_i^{\circ_C \circ_P} ,$$ where $L_0=E$, can be applied. Thus, we have
\begin{center}
$L_1=\Sigma^*\backslash (\Sigma^* \backslash (L(G) \backslash ((L(G)\backslash E)/\Sigma_{uc}^*)))\Sigma^*$
\end{center}
by straightforward applications of interior constructions. We can simplify the expression of $L_1$ as follows.
\begin{lemma}
\label{lemma: simple}
$L_1=L(G)\backslash ((L(G)\backslash E)/\Sigma_{uc}^*)\Sigma^*$.
\end{lemma}
\emph{Proof}: We now use the rule $A\backslash (B \backslash C)=(A\backslash B) \cup (A \cap C)$ and obtain the following 
\begin{center}
    $L_1=\Sigma^*\backslash (L(G)^c \cup (L(G)\backslash E)/\Sigma_{uc}^*)\Sigma^*=\Sigma^*\backslash (L(G)^c\Sigma^* \cup ((L(G)\backslash E)/\Sigma_{uc}^*)\Sigma^*)=(\Sigma^*\backslash L(G)^c\Sigma^*)\cap \Sigma^*\backslash (L(G)\backslash E)/\Sigma_{uc}^*)\Sigma^*$
\end{center}
We then apply the rule
$\Sigma^*\backslash L(G)^c\Sigma^*= L(G)$ and obtain the simplified expression $L_1=L(G)\backslash ((L(G)\backslash E)/\Sigma_{uc}^*)\Sigma^*$.
\hfill $\blacksquare$

$L_1$ is by construction prefix-closed, i.e., $L_1^{\circ_P}=L_1$. The next result shows that $L_1$ is indeed the fixed-point, i.e., $L_1 ={\bf Sup}\overline{\mathcal{C}}(E;L(G))$. 
\begin{lemma}
\label{lemma:fixedpoint}
$L_1^{\circ_C}=L_1$ or, equivalently, $L_1\Sigma_{uc}^*\cap L(G)=L_1$.
\end{lemma}
\emph{Proof}:
We need to show  $L_1\Sigma_{uc}^*\cap L(G) \subseteq L_1$. We only need to show
\begin{center}
    $(L(G)\backslash ((L(G)\backslash E)/\Sigma_{uc}^*)\Sigma^*)\Sigma_{uc}^*\cap L(G)  \cap ((L(G)\backslash E)/\Sigma_{uc}^*)\Sigma^* =\varnothing$
\end{center}
Suppose, on the contrary, that there exists some string $s \in L(G)\backslash ((L(G)\backslash E)/\Sigma_{uc}^*)\Sigma^*$ and some string $s_u \in \Sigma_{uc}^*$ such that $ss_u\in ((L(G)\backslash E)/\Sigma_{uc}^*)\Sigma^* \cap L(G)$.


We analyze $s$ based on $ss_u \in ((L(G)\backslash E)/\Sigma_{uc}^*)\Sigma^*$. There are only two (not necessarily mutually exclusive) cases.
\begin{enumerate}
    \item $s \in ((L(G)\backslash E)/\Sigma_{uc}^*)\Sigma^*$
    \item $s \in \overline{(L(G)\backslash E)/\Sigma_{uc}^*}$, and there exist some $s_u^1, s_u^2 \in \Sigma_{uc}^*$ such that $s_u=s_u^1s_u^2$ and $ss_u^1\in (L(G)\backslash E)/\Sigma_{uc}^*$.
\end{enumerate}
The first case is in contradiction with the supposition that $s \in L(G)\backslash ((L(G)\backslash E)/\Sigma_{uc}^*)\Sigma^*$. Thus, we only need to consider the second case that $s \in \overline{(L(G)\backslash E)/\Sigma_{uc}^*}$. 

However, from $ss_u^1\in (L(G)\backslash E)/\Sigma_{uc}^*$, we know that $s\in (L(G)\backslash E)/\Sigma_{uc}^*$, which again leads to contradiction. \hfill $\blacksquare$\\

Again, the formula $${\bf Sup}\overline{\mathcal{C}}(E;L(G))=L(G)\backslash ((L(G)\backslash E)/\Sigma_{uc}^*)\Sigma^*$$ is equivalent to the formula   derived in~\cite{BGRLMW90}. In particular, we have
\begin{center}${\bf Sup}\overline{\mathcal{C}}(E;L(G))=(E \cup (L(G)\backslash E))\backslash ((L(G)\backslash E)/\Sigma_{uc}^*)\Sigma^*=E\backslash ((L(G)\backslash E)/\Sigma_{uc}^*) $
\end{center}
\subsubsection{Topology Optimization for  Prefix-closed Controllability}
\label{sec: topop}
In certain cases, a system of topologized equations can be reduced to an equivalent topologized equation by setting up an  optimized semi-topology. It follows that the technique of Section~\ref{sec: atopo} can be used for solving such a system of topologized equations as well, as an acceleration technique (for chain termination). 

It is possible to obtain an optimized semi-topology for the computation of ${\bf Sup}\overline{\mathcal{C}}(E;L(G))$ as follows.

With $K \subseteq E$, we have
\begin{center}
$K=\overline{K}$ and $K\Sigma_{uc}^* \cap L(G)=K$ iff
$\overline{K}\Sigma_{uc}^* \cap L(G)=K$ iff $\overline{K}\Sigma_{uc}^* \cap (L(G)\backslash K)=\varnothing$ iff $\overline{K}\cap (L(G)\backslash K)/\Sigma_{uc}^*=\varnothing$ iff $K \cap ((L(G)\backslash K)/\Sigma_{uc}^*)\Sigma^*=\varnothing$ iff $K \subseteq L(G)\backslash ((L(G)\backslash K)/\Sigma_{uc}^*)\Sigma^*$ iff $K = L(G)\backslash ((L(G)\backslash K)/\Sigma_{uc}^*)\Sigma^*$ iff $K=L(G)\backslash (L(G)\backslash K)^{\square_O}$,
\end{center}
where $\square_O:2^{L(G)} \longrightarrow 2^{L(G)}$ is a map defined such that, for any $L \subseteq L(G)$, $L^{\square_O}:=(L/\Sigma_{uc}^*)\Sigma^* \cap L(G)$. $\square_O$ is a (semi)-topological closure operator on $L(G)$.

Thus, ${\bf Sup}\overline{\mathcal{C}}(E;L(G))$ exists in $\overline{\mathcal{C}}(E;L(G))$ and 
\begin{center}${\bf Sup}\overline{\mathcal{C}}(E;L(G))=L(G)\backslash (L(G)\backslash E)^{\square_O}=L(G)\backslash ((L(G)\backslash E)/\Sigma_{uc}^*)\Sigma^*$
\end{center}
\subsection{Computation With A System of Topologized Equations With Search Space Relaxation}
In this subsection, we show that the computation of the supremal controllable sublanguage, without the prefix-closedness property, can be treated as solving a system of topologized equations with search space relaxation. 

Let
\begin{center}
$\mathcal{C}(E; L(G))=\{K \subseteq E \mid \overline{K}\Sigma_{uc}^* \cap L(G) = \overline{K}\}$. 
\end{center}
denote the collection of controllable sublanguages of $E$ w.r.t. $(L(G), \Sigma_{uc})$. Without loss of generality, we assume $\overline{E} \subseteq L(G)$. The elements of $\mathcal{C}(E; L(G))$ can be reformulated as below (see Section~\ref{sec: to}): with $\overline{E} \subseteq L(G)$, 
\begin{center}
    $\overline{K}\Sigma_{uc}^* \cap L(G) = \overline{K}$ iff $\overline{K}= L(G) \backslash ((L(G)\backslash \overline{K})/\Sigma_{uc}^*)$ iff $\overline{K}=L(G)\backslash (L(G)\backslash \overline{K})^{\square_C}$
\end{center}

We can now apply the technique from Section~\ref{sec: relax}.  Thus, the iteration scheme $K_{i+1}=K_i \cap X_i$ can be used. $X_i$ is 
the supremal solution of the system 
\begin{align*}
X=\Sigma^*\backslash (\Sigma^*\backslash  X)^{ \square_P}, \\
X=L(G)\backslash (L(G)\backslash X)^{\square_C},
\end{align*}
where $X \subseteq \overline{K_i}$. Thus,
$X_i=L(G)\backslash ((L(G)\backslash \overline{K_i})/\Sigma_{uc}^*)\Sigma^*$, by reusing the result from Section~\ref{sec: pc}. We then have the iteration equation
\begin{center}
$K_{i+1}=K_i \cap X_i=K_i\backslash ((L(G)\backslash \overline{K_i})/\Sigma_{uc}^*)\Sigma^*$
\end{center}
\subsection{Computation With A Mixed System of Topologized Equations}
In this subsection, we explain the application of a mixed system of topologized equations to the computation of the supremal controllable and normal sublanguage~\cite{WC18}.

Let
\begin{center}
$\overline{N}(E; L(G))=\{K \subseteq E \mid \overline{K}=P_{\Sigma_o}^{-1}P_{\Sigma_o}(\overline{K}) \cap L(G)\}$
\end{center}
denote the collection of sublanguages of $E$ whose prefix-closure is normal w.r.t. $(L(G), \Sigma_o)$. Now, let $$\mathcal{S}(E)=\mathcal{C}(E; L(G))\cap N(E; L_m(G)) \cap \overline{\mathcal{N}}(E; L(G)).$$ $\mathcal{S}(E)$ is commonly referred to as the set of controllable and normal sublanguages of $E$~\cite{WC18},~\cite{ZMW05}. Now, without loss of generality, we shall assume $E \subseteq L_m(G)$ and $\Sigma_c \subseteq \Sigma_o$. 
The elements of $\mathcal{S}(E)$ are formulated as the solutions of the system
\begin{align*}
\overline{K}=L(G)\backslash (L(G)\backslash \overline{K})^{\square_O}, \\
K=L_m(G) \backslash (L_m(G)\backslash K)^{\square_N},\\
\overline{K}=L(G) \backslash (L(G)\backslash \overline{K})^{\square_N}
\end{align*}
with $K \subseteq E$. Then, we can apply the technique developed in Section~\ref{sec: mixed}. Here, we directly use an acceleration technique, with an optimized topology. The key result is the following~\cite{WL18}.
\begin{lemma}
Let $K \subseteq E 
\subseteq L_m(G)$. We have that
$\overline{K}=P_{\Sigma_o}^{-1}P_{\Sigma_o}(\overline{K}) \cap L(G)$ and $\overline{K}\Sigma_{uc}^* \cap L(G) = \overline{K}$ if and only if $\overline{K}=P_{\Sigma_o}^{-1}P_{\Sigma_o}(\overline{K})\Sigma_{uc}^* \cap L(G)$.
\end{lemma}
With $K \subseteq E $, we have 
\begin{center}
    $\overline{K}=P_{\Sigma_o}^{-1}P_{\Sigma_o}(\overline{K})\Sigma_{uc}^* \cap L(G)$ iff $P_{\Sigma_o}^{-1}P_{\Sigma_o}(\overline{K})\Sigma_{uc}^* \cap L(G) \subseteq \overline{K}$ iff $P_{\Sigma_o}^{-1}P_{\Sigma_o}(\overline{K})\Sigma_{uc}^* \cap (L(G) \backslash \overline{K})=\varnothing$ iff  $P_{\Sigma_o}^{-1}P_{\Sigma_o}(\overline{K}) \cap (L(G) \backslash \overline{K})/\Sigma_{uc}^*=\varnothing$ iff $\overline{K} \cap P_{\Sigma_o}^{-1}P_{\Sigma_o}((L(G) \backslash \overline{K})/\Sigma_{uc}^*)=\varnothing$ iff $\overline{K} \subseteq L(G)\backslash  P_{\Sigma_o}^{-1}P_{\Sigma_o}((L(G) \backslash \overline{K})/\Sigma_{uc}^*)$ iff $\overline{K} = L(G)\backslash  P_{\Sigma_o}^{-1}P_{\Sigma_o}((L(G) \backslash \overline{K})/\Sigma_{uc}^*)$
\end{center}
Let $\square_A: 2^{L(G)}\longrightarrow 2^{L(G)}$ denote a function defined such that $L^{\square_A}=P_{\Sigma_o}^{-1}P_{\Sigma_o}(L/\Sigma_{uc}^*) \cap L(G)$ for any $L \subseteq L(G)$. $\square_A$ is a (semi)-topological closure operator over $L(G)$, with $\Sigma_c \subseteq \Sigma_o$. 

Thus, the elements of $\mathcal{S}(E)$ can be reformulated as the solutions of the system
\begin{align*}
\overline{K}=L(G)\backslash (L(G)\backslash \overline{K})^{\square_A}, \\
K=L_m(G) \backslash (L_m(G)\backslash K)^{\square_N},
\end{align*}
with $K \subseteq E$. We now apply the technique of Section~\ref{sec: mixed} to this system, which leads to the  iteration scheme
\begin{center}
     $K_{j+1}=K_j^{\diamond_A\diamond_N}=K_j^{\diamond_A\circ_N}=L_m(G)\backslash P_{\Sigma_o}^{-1}P_{\Sigma_o}(L_m(G)\backslash K_j^{\diamond_A})$, with $K_0=E$
\end{center}
In particular, $K_j^{\diamond_A}$ is the supremal solution of the equation $\overline{K}=L(G)\backslash (L(G)\backslash \overline{K})^{\square_A}$, with $K \subseteq K_j$. It
can be solved with the iteration scheme (see   Section~\ref{sec: relax}).
\begin{center}
$L_{ i+1}=L_i \cap X_i$, where $L_0=K_j$
\end{center}where $X_i$ is the supremal solution of the system 
\begin{align*}
X=\Sigma^*\backslash (\Sigma^*\backslash X)^{\square_P}, \\
X=L(G) \backslash (L(G)\backslash X)^{\square_A},
\end{align*}
with $X \subseteq \overline{L_i}$. $X_i$ can be computed with the  scheme
\begin{center}$H_{k+1}=H_k^{\circ_A \circ_P}$, where $H_0=\overline{L_i}$
\end{center}
We have 
\begin{center}
    $H_1=\Sigma^*\backslash (\Sigma^*\backslash (L(G)\backslash P_{\Sigma_o}^{-1}P_{\Sigma_o}((L(G)\backslash \overline{L_i})/\Sigma_{uc}^*)))\Sigma^*=\Sigma^*\backslash (L(G)^c \cup P_{\Sigma_o}^{-1}P_{\Sigma_o}((L(G)\backslash \overline{L_i})/\Sigma_{uc}^*))\Sigma^*=\Sigma^*\backslash (L(G)^c\Sigma^* \cup P_{\Sigma_o}^{-1}P_{\Sigma_o}((L(G)\backslash \overline{L_i})/\Sigma_{uc}^*)\Sigma^*)=L(G)\backslash P_{\Sigma_o}^{-1}P_{\Sigma_o}((L(G)\backslash \overline{L_i})/\Sigma_{uc}^*)\Sigma^*=L(G)\backslash P_{\Sigma_o}^{-1}P_{\Sigma_o}(((L(G)\backslash \overline{L_i})/\Sigma_{uc}^*)\Sigma^*)$
\end{center}
$H_1$ is by construction prefix-closed, i.e., $H_1^{\circ_P}=H_1$. The next result shows that $H_1$ is indeed the fixed-point, i.e., $H_1 =H_1^{\circ_A \circ_P}$. 
\begin{lemma}
\label{lemma:fixedpoint}
$H_1^{\circ_A}=H_1$ or, equivalently, $P_{\Sigma_o}^{-1}P_{\Sigma_o}(H_1)\Sigma_{uc}^*\cap L(G)=H_1$.
\end{lemma}
\emph{Proof}: We need to show $P_{\Sigma_o}^{-1}P_{\Sigma_o}(H_1)\Sigma_{uc}^*\cap L(G) \subseteq H_1$. We only need to show 
\begin{center}
    $P_{\Sigma_o}^{-1}P_{\Sigma_o}(L(G)\backslash P_{\Sigma_o}^{-1}P_{\Sigma_o}(((L(G)\backslash \overline{L_i})/\Sigma_{uc}^*)\Sigma^*))\Sigma_{uc}^* \cap L(G) \cap P_{\Sigma_o}^{-1}P_{\Sigma_o}(((L(G)\backslash \overline{L_i})/\Sigma_{uc}^*)\Sigma^*)=\varnothing$.
\end{center}
Suppose, on the contrary, that there exists some string $$s \in P_{\Sigma_o}^{-1}P_{\Sigma_o}(L(G)\backslash P_{\Sigma_o}^{-1}P_{\Sigma_o}(((L(G)\backslash \overline{L_i})/\Sigma_{uc}^*)\Sigma^*))$$ and some string $s_u \in \Sigma_{uc}^*$, such that $$ss_u \in L(G) \cap P_{\Sigma_o}^{-1}P_{\Sigma_o}(((L(G)\backslash \overline{L_i})/\Sigma_{uc}^*)\Sigma^*).$$
We analyze $s$ based on $ss_u \in  P_{\Sigma_o}^{-1}P_{\Sigma_o}((L(G)\backslash \overline{L_i})/\Sigma_{uc}^*)\Sigma^*$. There are two (not necessarily mutually exclusive) cases.
\begin{enumerate}
    \item 
    $s \in P_{\Sigma_o}^{-1}P_{\Sigma_o}((L(G)\backslash \overline{L_i})/\Sigma_{uc}^*)\Sigma^*$
    \item $s \in \overline{P_{\Sigma_o}^{-1}P_{\Sigma_o}((L(G)\backslash \overline{L_i})/\Sigma_{uc}^*)}$, and also there exist some $s_u^1, s_u^2 \in \Sigma_{uc}^*$ such that $s_u=s_u^1s_u^2$ and $ss_u^1\in P_{\Sigma_o}^{-1}P_{\Sigma_o}((L(G)\backslash \overline{L_i})/\Sigma_{uc}^*)$.
\end{enumerate}
The first case is in contradiction with $$s \in P_{\Sigma_o}^{-1}P_{\Sigma_o}(L(G)\backslash P_{\Sigma_o}^{-1}P_{\Sigma_o}(((L(G)\backslash \overline{L_i})/\Sigma_{uc}^*)\Sigma^*)).$$ since (with the rule $P_{\Sigma_o}^{-1}P_{\Sigma_o}(A \backslash P_{\Sigma_o}^{-1}P_{\Sigma_o}(B))\cap P_{\Sigma_o}^{-1}P_{\Sigma_o}(B)=\varnothing$)
\begin{center}
    $P_{\Sigma_o}^{-1}P_{\Sigma_o}(L(G)\backslash P_{\Sigma_o}^{-1}P_{\Sigma_o}(((L(G)\backslash \overline{L_i})/\Sigma_{uc}^*)\Sigma^*)) \cap P_{\Sigma_o}^{-1}P_{\Sigma_o}(((L(G)\backslash \overline{L_i})/\Sigma_{uc}^*)\Sigma^*)=\varnothing$.
\end{center}
Thus, we only need to consider the second case. However, from $ss_u^1\in P_{\Sigma_o}^{-1}P_{\Sigma_o}((L(G)\backslash \overline{L_i})/\Sigma_{uc}^*)$, we conclude, with $\Sigma_c \subseteq \Sigma_o$, that
\begin{center}
$s \in P_{\Sigma_o}^{-1}P_{\Sigma_o}((L(G)\backslash \overline{L_i})/\Sigma_{uc}^*)/\Sigma_{uc}^*=P_{\Sigma_o}^{-1}P_{\Sigma_o}((L(G)\backslash \overline{L_i})/\Sigma_{uc}^*)$.
\end{center}
However, this again contradicts with
\begin{center}$s \in P_{\Sigma_o}^{-1}P_{\Sigma_o}(L(G)\backslash P_{\Sigma_o}^{-1}P_{\Sigma_o}(((L(G)\backslash \overline{L_i})/\Sigma_{uc}^*)\Sigma^*))\subseteq P_{\Sigma_o}^{-1}P_{\Sigma_o}(L(G)\backslash P_{\Sigma_o}^{-1}P_{\Sigma_o}((L(G)\backslash \overline{L_i})/\Sigma_{uc}^*)).$
\end{center}
\hfill $\blacksquare$ \\
Thus, we have $L_{i+1}=L_i\backslash P_{\Sigma_o}^{-1}P_{\Sigma_o}((L(G)\backslash \overline{L_i})/\Sigma_{uc}^*)\Sigma^*$.

In summary, we have the iteration equation
\begin{center}$K_{j+1}=L_m(G)\backslash P_{\Sigma_o}^{-1}P_{\Sigma_o}(L_m(G)\backslash K_j^{\diamond_A})$, with $K_0=E$.
\end{center}$K_j^{\diamond_A}$ is computed with the iteration equation \begin{center}$L_{i+1}=L_i\backslash P_{\Sigma_o}^{-1}P_{\Sigma_o}((L(G)\backslash \overline{L_i})/\Sigma_{uc}^*)\Sigma^*$, with $L_0=K_j$
\end{center}

\section{Conclusion}
\label{sec: Con}
In this work, we have presented a topological technique for computing some supremal sublanguages, often specified by language equations, that arise from the study of the supervisory control theory. Theorem~\ref{them: supremal} of Section~\ref{sec: atopo} is central to the rough set theory~\cite{P92}, which is a formal theory of approximation. This work can thus be viewed as a further development of the rough set theory in solving language equations. 

The topological technique provides an elementary and uniform approach for the computation, if we do not care about the finite convergence property. It is of interest to know if the topological perspective can also be useful for establishing the finite convergence property, which is left open. 

\end{document}